\documentclass[alpha-refs]{article}
\usepackage{siunitx}
\usepackage{subfigure}
\usepackage{booktabs}
\usepackage{tikz}
\usetikzlibrary{patterns}
\usepackage{pgfplotstable}  	
\usepackage{import}
\usepackage{authblk}
\usepackage{natbib}
\usepackage{amssymb,amsmath}
\usepackage[margin=1in]{geometry}
\usepackage{caption}

\usepackage[left]{lineno}
\title{Scale interactions and anisotropy in stable boundary layers}

\author[1]{Nikki Vercauteren}
\author[1]{Vyacheslav Boyko}
\author[2,3]{Davide Faranda}
\author[4]{Ivana Stiperski}

\affil[1]{FB Mathematik und Informatik, Freie Universit{\"a}t Berlin, 14195 Berlin, Germany}
\affil[2]{LSCE-IPSL, CEA Saclay l'Orme des Merisiers, CNRS UMR 8212 
CEA-CNRS-UVSQ, Universit\'e Paris-Saclay, 91191 Gif-sur-Yvette, France}
\affil[3]{London Mathematical Laboratory, London, United Kingdom}
\affil[4]{Institute of Atmospheric and Cryospheric Sciences, University of Innsbruck, Innsbruck, Austria}

\begin{document}

\maketitle

\begin{abstract}	
Regimes of interactions between motions on different time-scales are 
investigated in the FLOSSII dataset for nocturnal near-surface stable boundary 
layer (SBL) turbulence. The non-stationary response of turbulent vertical 
velocity variance to non-turbulent, sub-mesoscale wind velocity variability is 
analysed using the bounded variation, finite element, vector autoregressive 
factor models (FEM-BV-VARX) clustering method. Several locally stationary flow 
regimes are identified with different influences of sub-meso wind velocity on 
the turbulent vertical velocity variance. In each flow regime, we analyse 
multiple scale interactions and quantify the amount of turbulent variability 
which can be statistically explained by external forcing by the sub-meso wind velocity. The 
state of anisotropy of the Reynolds stress tensor in the different flow regimes 
is shown to relate to these different signatures of scale interactions. In flow 
regimes under considerable influence of the sub-mesoscale wind variability, the Reynolds stresses show 
a clear preference for strongly anisotropic, one-component states. These periods additionally show stronger persistence in their 
dynamics, compared to periods of more isotropic stresses. The analyses give 
insights on how the different topologies relate to non-stationary turbulence 
triggering by sub-mesoscale motions.
%
% Please include a maximum of seven keywords
%\keywords{statistical clustering, regime detection, non-stationary turbulence, nocturnal boundary layer, submeso motions, anisotropy tensor, persistent dynamics}
\end{abstract}
\section{Introduction}
Nocturnal and stable boundary layers (SBL) represent a challenge to numerical 
weather prediction \citep{Sandu:2013ia, Holtslag:2013iy}. Difficulties arise 
due to the unsteady nature of the flow caused by the interactions of processes 
on multiple scales. In strong wind conditions, mechanical forcing of turbulence 
is enough to overcome buoyant damping and turbulence is generally continuous 
and rather well described by classical similarity theory. In weak-wind 
conditions however, sporadic turbulence can be triggered by localised shear 
accelerations due to sub-mesoscale motions such as internal gravity waves, 
density currents, wind gusts or other motions \citep{Sun:2004tj, 
Mahrt:2014wc, Sun:2015km,Mortarini:2017em}. The non-turbulent, small-scale 
motions take a variety of forms and are poorly understood and not represented 
in models \citep{Belusic:2010jc, Kang:2015hi, Lang:2017bu}. Such non-local 
scale interactions modify the characteristics of boundary layer turbulence, 
giving it an intermittent nature, with a tendency to be decoupled from the 
surface \citep{Acevedo:2015iw}. In this very stable regime, classical 
surface-based parameterisations of turbulence fail at representing turbulence 
resulting from interactions with non-turbulent unsteady flow accelerations.\par 
Non-stationary turbulence under the influence of submeso motions has been 
analysed with different data analysis methods \citep{Mahrt:2012fs, 
Mahrt:2015es, Cava:2016ho, Mortarini:2017em}. A method to classify flow 
regimes based on the influence of submeso motions has been proposed recently 
using non-stationary vector autoregessive factor models with external influences (VARX) 
and data clustering based on a finite element, bounded variation method (FEM-BV-VARX) \citep{Horenko:2010ce}. 
The methodology provides the means 
of objectively classifying non-stationary dynamics influenced by external 
variables. This statistical clustering technique has proven powerful for 
classifying large scale atmospheric flow data, e.g. for identifying global 
atmospheric circulation regimes and blockings \citep{OKane:2016fd, 
Risbey:2015gh}. In the case of SBL turbulence, a combination of multiscale data 
filtering and FEM-BV-VARX clustering, was used by \cite{Vercauteren:2015fq} and 
\cite{Vercauteren:2016kx} to characterise the interactions between 
sub-mesoscale non-turbulent motions and turbulence. This strategy was found to 
successfully identify periods in which non-turbulent motions are active and statistically modulate the 
turbulence dynamics, corresponding to very stable periods.\par
As an additional effect on the structure of turbulence, stable temperature 
stratification results in a strong attenuation of the vertical turbulent 
motions by buoyancy forces, while shear forcing exerts a straining action. The 
combination of these effects can lead to strongly anisotropic turbulence 
\citep{Smyth:2000du}. Turbulence anisotropy leads to additional difficulties in 
parameterising turbulence, and \cite{Stiperski:2017db} have recently given a 
new perspective on the failure of traditional similarity scaling by relating it 
to the topology of the Reynolds stress tensor, based on observational evidence. 
Anisotropy of the energy-containing scales is quantified using the anisotropy 
stress tensor and the structure of the tensor can be conveniently represented 
using two invariants \citep{LUMLEY1979123}. Based on the eigenvalues of the 
anisotropy tensor, one can classify turbulence according to three limiting 
states \citep{pope2000turbulent}. The one-component limiting state describes a 
flow where one eigenvalue is much larger than the other two (sometimes 
referred to as rod-like turbulence), while the two-component limit has two 
directions with equal magnitude (and is sometimes referred to as pancake-like 
turbulence) and all three directions have equal magnitude in the three 
component, isotropic limit. \cite{Stiperski:2017db} showed that while close to 
isotropic and close to two-component axisymmetric stresses agreed well with 
existing SBL scaling relationships, one-component axisymmetric stresses 
deviated strongly from similarity scaling. Delineating isotropic from one or 
two-component cases appeared possible by using a combination of the wind speed 
and the turbulent kinetic energy (TKE), but differentiating between the one- 
and two-component cases themselves appeared more difficult. The delineation 
based on the wind speed and TKE is the approach taken by the Hockey Stick 
Transition (HOST) framework \citep{Sun:2015km}, based on the observation that 
turbulence shows two distinct behaviours depending on the wind speed. The low 
wind speed part of the HOST framework corresponds to intermittent turbulence 
where the TKE is quasi-invariant with the mean wind speed, and where the 
Reynolds stress tensor is highly anisotropic \citep{Stiperski:2017db}. This 
regime is also characterised by the presence of myriads of anisotropic 
sub-mesoscale motions which act as trigger for turbulence. The combination of 
multiple scale analysis of turbulence and statistical clustering proposed by 
\cite{Vercauteren:2015fq} and \cite{Vercauteren:2016kx} allows to identify flow 
regimes in which scales interact differently, and the regimes may correspond to different 
states of anisotropy.\par
The forcing of turbulence by wave-like non-turbulent motions typically occurs 
on scales just above the turbulent scales. In the Kolmogorov view of 
turbulence, the cascade of energy from large to small scales is accompanied by 
a loss of information about the geometry of the large scales. The unsteady 
forcing may result in non-equilibrated, highly anisotropic turbulence as 
indicated by multiscale decompositions in \cite{Vercauteren:2016kx}. According 
to the theory, highly anisotropic turbulence should tend to equilibrate to 
quasi-isotropic states at the small scales \citep{pope2000turbulent}. The 
dynamical evolution of large scale anisotropic structures towards isotropy is 
however a subject of research, which can be studied based on the anisotropy 
tensor structure. \cite{KWINGSOCHOI:2018hj} showed experimental evidence that 
the rate of return to isotropy depends on the initial topological state and is 
very slow for cigar-shaped, axisymmetric turbulence at a high Reynolds number. 
Their analyses also showed that the rate of return to isotropy was 
not linearly proportional to the degree of anisotropy as assumed by Rotta's classical return-to-isotropy model 
(see eg. \cite{pope2000turbulent}), but followed more complex, nonlinear dependences on the anisotropy tensor. \cite{Brugger:2018go} analysed the route to isotropy based on 
atmospheric measurements in the surface layer for canopy flows and highlighted 
a large influence of thermal stratification. Their analyses showed that 
trajectories in the phase space (for decreasing scales) defined by the 
anisotropy invariants deviate from those of return-to-isotropy known for 
homogeneous turbulence. In stably stratified conditions, the influence of 
anisotropic sub-mesoscale motions probably affects the turbulence anisotropy 
dynamics and this topic is investigated here.\par
A way to analyse the dynamical evolution of turbulence depending on its 
initial topology is to consider the dynamics in the anisotropy phase space 
defined by the invariants of the anisotropy tensor. \cite{Lucarini:2016ug} 
recently developed indicators that proved able to quantify the persistence of 
dynamics in phase space as well as the local dimension of the dynamics and 
applied them in the context of climate dynamics \citep{Faranda:2017bi, 
messori2017dynamical}. Quantifying the persistence of 
the turbulent states depending on their topology informs us on how 
constrained or not the dynamical evolution of turbulence is in a given starting 
state of anisotropy. Quantifying the dimension of the dynamics at each point of 
the anisotropy phase space additionally enables us to investigate the 
existence of preferred directions of the evolution of turbulent 
states of anisotropy.\par
In this work we will focus on following questions: 
Do flow regimes separated according to their scale-interactions properties 
correspond to unique states of anisotropy? How persistent are different states 
of anisotropy and is there a preferred trajectory in the anisotropy phase space 
in stably stratified conditions? We will address these questions based on turbulence 
measurements from the Fluxes Over Snow Surfaces II campaign (FLOSSII).\par
\section{Methods}
\subsection{Dataset}
The analysis is based on turbulence data collected during the Fluxes Over Snow 
Surfacess II (FLOSSII) experiment that was conducted  from 20 November 2002 to 
4 April 2003 over a locally flat grass surface south of Walden, Colorado, USA, 
in the Arapaho National Wildlife Refuge \citep{Mahrt:2005et}. The surface was 
often covered by a thin snow layer during the field program. The turbulence is 
measured by a tower collecting data at 1, 2, 5, 10, 15, 20 and 30 m with 
Campbell CSAT3 sonic anemometers.  The data set was quality controlled and 
segments of instrument problems and meteorologically impossible values were 
eliminated (Larry Mahrt, personal communication).\par
The following analysis is based on night-time data, taken between 18:00 and 
7:00, Local Standard Time. The period is selected because the surface sensible 
heat flux averaged over all nights is negative during this time. Flow regime 
identification based on the FEM-BV-VARX clustering methodology (see Section 
\ref{Sec:clustering}) ideally requires continuous data, however the dataset 
will consist of continuous night-time data separated by gaps during the day. In 
order to maximise continuity of the dataset, nights with data gaps 
longer than 80 minutes (12 nights) as well as nights with wind 
flowing persistently  through the measurement tower for periods longer than 5 
minutes (51 nights) were removed from the analysis. The resulting 68 nights 
left for analysis have data gaps shorter than 1 minute and are deemed mostly 
uncontaminated. The short gaps are linearly interpolated. The 60Hz raw data are 
rotated into the mean wind direction based on 30  minutes averages using Double 
Rotation.\par
The flow characteristics of the FLOSSII dataset were analysed by 
\cite{Mahrt:2011bz}, showing complex non-stationary relationships between 
turbulence and sub-mesoscale wind velocity that will be analysed with the 
objective FEM-BV-VARX classification strategy here. Analyses of non-turbulent 
structures identified as sub-mesoscale motions by the Turbulent Event Detection 
(TED) method \citep{Kang:2015hi} revealed the presence of complex structures 
affecting the turbulent dynamics.\par
\subsection{Extracting scales of motion}\label{Sec:filter}
The multiresolution flux decomposition (MRD) \citep{Vickers:2003tw} and other 
wavelet analysis tools have been successfully used to analyse SBL scale-wise 
properties of flux. The MRD can be used to assess the amount of flux that is 
due to 
eddies of a certain size, thereby providing a way to identify a cospectral gap 
scale. The gap scale is usually identified as the scale at which the flux 
crosses the zero-line and indicates the appropriate averaging period needed to 
separate 
contributions of non-turbulent sub-mesoscale of motion from turbulent 
fluxes. The MRD analyses of the nocturnal FLOSSII data show that the cospectral 
gap scale depends on the flow regime (not shown), but that scales 
smaller than approximately 1 minute mainly correspond to turbulent fluxes. We 
therefore define the turbulent vertical velocity fluctuations as $\sigma_w = 
\sqrt{\overline{w'w'}}$, where $w$ is the vertical wind velocity component, the 
overbar denotes an averaging period of 1 minute and the prime denotes 
deviations from the average.  A sub-mesoscale horizontal mean wind speed is defined as:
\begin{equation}\label{Eq:Vsmeso}
	V_{smeso} =  \sqrt{u_s^2 + v_s^2}\, ,
\end{equation} 
where $u_s$ and $v_s$ are the streamwise and lateral velocity components on sub-mesoscales. In the 
definition of the sub-mesoscale 
fluctuations $\phi_s = \overline{\phi} - [\phi]$, the overbar denotes a 1-min 
averaging time and the square brackets denote a 30-min averaging time, such 
that these fluctuations represent the deviations of the 1-min sub-record 
averages from the 30-min average. These definitions of turbulent vertical 
velocity fluctuations and sub-mesoscale wind velocity fluctuations are used to 
analyse the non-stationary interactions between the submeso velocity scale and turbulence. 
This choice of timescales is identical to the choice made by 
\cite{Vercauteren:2015fq} and \cite{Vercauteren:2016kx} to analyse scale 
interactions in the SnoHATS dataset of SBL turbulence.\par
\subsection{Clustering flow regimes}\label{Sec:clustering}
The Finite Element, Bounded Variation, Vector Auto-Regressive with eXternal 
factors (FEM-BV-VARX) method \citep{Horenko:2010gu, Horenko:2010ce} relates an 
observed variable of interest at a discrete-time $t$ ( in our context $\sigma_w(t)$) to the past history 
of observations ($\sigma_w(t-p\tau)$, where $p \in \mathbb{Z}^{>0}$ and $\tau$ is the 
discrete-time unit step equal to the inverse sampling frequency). Influences from external forcing variables can also be considered. In order to cluster different stable flow regimes, we start with the hypothesis that in some flow regimes the turbulence may be modulated to a large extent by sub-mesoscale motions. These are expected to correspond to weak-wind, very stable periods. We therefore consider extracted sub-mesoscale horizontal velocity as the external forcing variable influencing turbulence in the statistical model. Our classification goal is to separate cases during which the time evolution of turbulence is modulated by the time evolution of the sub-mesoscale wind velocity from cases during which the response of turbulence to forcing by sub-mesoscales is different or less apparent. In order to achieve this goal using the FEM-BV-VARX method, we assume that the evolution in time of the turbulent 
mixing can be approximated by a locally stationary statistical process (VARX) 
that is influenced by the horizontal wind speed at specified sub-mesoscale 
motions. The method is thus applied to identify 
different SBL flow regimes, based on the interactions of different scales of 
motion. The VARX model relates the dynamics of 
$\sigma_w(t)$ to the external factors 
$V_{smeso}(t-p\tau)$, and the relationship is modulated by a set of time-dependent, piecewise constant 
parameters $\overline{\Theta}(t)$. The VARX model in our application takes the following form:
\begin{equation}\label{Eq:VARX}
\sigma_{w}(t) = \mu (t) +B_{0}(t)V_{smeso}(t) + B_{1}(t)V_{smeso}(t-\tau) + 
\cdots +B_{p}(t)V_{smeso}(t-p\tau) +C(t)\varepsilon(t)\, ,
\end{equation}
where the process $\sigma_{w}(t)$ is the time evolution of the vertical 
velocity variance measured at one location; the external factor is the time 
evolution of the streamwise velocity on our previously defined sub-mesoscales 
$V_{smeso}$ (Eq. \ref{Eq:Vsmeso}). $\varepsilon_{t}$ is a noise process with 
zero 
expectation, the parameters $\overline{\Theta}(t) = \left( \mu (t), B(t), C(t) 
\right)$ are time- dependent model coefficients for the VARX process and $p$ is the 
memory depth of the external factor. The model assumes a linear relationship 
between $\sigma_{w}$ and $V_{smeso}$, which was shown to be appropriate by 
\cite{Mahrt:2011bz}, based on the FLOSSII dataset and by 
\cite{Vercauteren:2015fq} for the SnoHATS dataset. Both analyses however 
highlighted that the linear dependence of $\sigma_{w}$ on the sub-mesoscale 
wind speed is not always constant. The turbulence relates to different scales 
of motions in a complex, non-stationary way and we use the FEM-BV-VARX 
clustering method to disentangle the different relationships. Since our 
interest lies mainly in characterising scale interactions, we do not consider 
an autoregressive part in the model (\ref{Eq:VARX}). The model assumes a number $K$ of statistical 
processes that corresponds to the number of clusters. The clustering method assumes 
that the dynamics is persistent in the following way: over time scales 
that are long compared with the characteristic fluctuation time scales in the data,
the timeseries of $\sigma_{w}$ is best represented by a VARX process with a fixed set of parameters.
Once in a while, however, the flow regime changes, and a different VARX provides a 
better reproduction of the time series' main characteristics. The FEM-BV-VARX method 
simultaneously detects such regime changes and estimates the optimal model coefficients for the VARX processes.
The assumption of local stationarity of the statistical process is enforced by setting a persistence 
parameter $C_{p}$, which defines the maximum allowed number of transitions 
between a total of $K$ different statistical processes. Each individual statistical 
process corresponds to a different set of constant values 
of the VARX model coefficients $\overline{\Theta}$, meaning that the coefficients $\overline{\Theta}(t)$ are locally constant in time, and change value at every transition between regimes. The cluster states are 
indicated by a cluster affiliation function, which is calculated by the procedure. The reader 
is referred to \cite{Horenko:2010ce} for information regarding the minimisation 
procedure used to solve the clustering problem. More detailed explanations on 
the application of the classification scheme to SBL turbulence is given in 
\cite{Vercauteren:2015fq}. User defined parameters and their choice are 
discussed in the results section \ref{Sec:VARX}.\par
\subsection{Anisotropy of the Reynolds stress tensor}
The forces imposed on the mean flow by the turbulent fluctuations are quantified using the Reynolds stress tensor 
\begin{equation}
\overline{u_{i}'u_{j}'} = \begin{pmatrix}
\overline{u_{1}'u_{1}'} & \overline{u_{1}'u_{2}'} & \overline{u_{1}'u_{3}'}  \\
\overline{u_{2}'u_{1}'} & \overline{u_{2}'u_{2}'} & \overline{u_{2}'u_{3}'}  \\
\overline{u_{3}'u_{1}'} & \overline{u_{3}'u_{2}'} & \overline{u_{3}'u_{3}'}  
\end{pmatrix},
\end{equation}
where $u_{i}'$ denotes velocity fluctuations and the overbar represents time average, i.e. $u_{i}'=u_{i}-\overline{u_{i}}$.
The anisotropic components of the tensor effectively transport momentum, while the isotropic or diagonal components can be absorbed in a modified mean pressure \citep{pope2000turbulent}. A way to characterise anisotropy of turbulence is to use symmetric, traceless tensors whose elements vanish in isotropic flows. The anisotropy tensor
\begin{equation}\label{Eq:anis}
a_{ij}= \frac{\overline{u_{i}u_{j}}}{2k}- \frac{\delta_{ij}}{3}, \quad k= \frac{\overline{u_{i}u_{i}}}{2},
\end{equation}
in which $\delta$ is the Kronecker delta and summation over repeated indices is implied, satisfies this condition. It was  introduced by \cite{LUMLEY1979123} to describe the evolution of turbulence towards isotropy in homogeneous, anisotropic flows. The anisotropy tensor has two independent principal scalar invariants ($II=a_{ij}a_{ji}$, $III=a_{ij}a_{in}a_{jn}$) which can be obtained from the eigenvalues of the tensor and are independent of the coordinate system. The invariants, or equivalently the eigenvalues of the tensor, are used to describe the relative strength of the fluctuating velocity components \citep{pope2000turbulent}. Mapping the values of the two invariants on the plane defined by $II$ and $III$ gives a simple graphical representation of the different states of anisotropy of the Reynolds stresses. This representation was initially proposed by \cite{Lumley:2006jy} and called an anisotropy invariant map. A functional relationship between $II$ and $III$ further defines a bounded region - the \emph{Lumley triangle} - on the plane of the invariants in which all physically realisable turbulence is found. Based on the position on that triangle, one can estimate the type of anisotropy of the Reynolds stress tensor. Indeed, the invariant $II$ quantifies the degree of anisotropy of the Reynolds stress, while the invariant $III$ encodes the topological characteristic of the anisotropy, with positive values indicating mostly one-component turbulence and negative values, mainly two-component axisymmetric turbulence. As such there are three special limiting states that correspond to edges in the Lumley triangle: the isotropic limit where all the eigenvalues are equal, the two-component limit where two eigenvalues are of equal magnitude and much large than the third eigenvalue and the one-component limit where one eigenvalue is much larger than the two others. \par
\begin{figure}[hbt]
	\centering
	\includegraphics{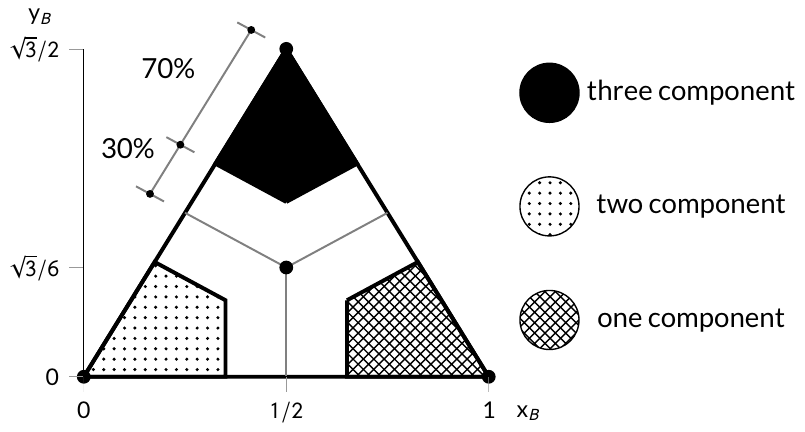}
	\caption{Definition of the anisotropy states in the barycentric map.}
	\label{Fig:BAM}
\end{figure}

As an equivalent alternative to the Lumley triangle, a barycentric Lumley map based on a linear domain that equally weighs the different limiting states of anisotropy simplifies the graphical interpretation of anisotropy of turbulence, avoiding nonlinear distortions \citep{Banerjee:2009jk}. The limiting states are placed at $x_{1C} = (1, 0)$, $ x_{2C} = (0, 0)$, and $x_{3C} = (1/2, \sqrt{3/2})$ and correspond respectively to purely one-component anisotropy (with one dominant eigenvalue), two-component axisymmetric anisotropy (with two dominant eigenvalues of equal magnitude) and to the three component, isotropic limit. Any anisotropy state is located as a point $(x_{B} , y_{B} )$ in this phase space such that the linear combination holds (see Fig. \ref{Fig:BAM})
\begin{align}
x_{B} &= C_{1C} x_{1C} +C_{2C} x_{2C} +C_{3C} x_{3C} = C_{1C} +\frac{1}{2}C_{3C}, \\
y_{B} &= C_{1C} y_{1C} +C_{2C} y_{2C} +C_{3C} y_{3C} = \frac{\sqrt{3}}{2}C_{3C}. 
\end{align}
The corresponding weights ($C_{1C},C_{2C},C_{3C}$) are entirely determined by 
the eigenvalues $\lambda_i$  ($i=1, 2, 3$)
of the normalised Reynolds stress anisotropy tensor, such that $C_{1C} = 
\lambda_1 - \lambda_2$, $C_{2C} = 2(\lambda_2 - \lambda_3)$, and $C_{3C} = 
3\lambda_3 + 1$. \par
Following \cite{Stiperski:2017db}, we define three regions in the barycentric 
Lumley map that correspond to anisotropy states close to each of the three pure 
limiting states. These regions are determined as kite-shaped regions of the 
barycentric map illustrated in Figure \ref{Fig:BAM}. The limiting lines for 
each kite were chosen to cover 70 \% of the sides of the equilateral triangle. 
Anisotropy states falling within each of the limiting regions will be denoted 
as \emph{pure anisotropy states}.\par
\subsection{Persistence and dimension of dynamical states of anisotropy} 
\label{Sec:persistency}
The Lumley triangle or its barycentric map counterpart define a geometric set that encapsulates all physically realisable turbulence states. 
In the mathematical field of dynamical systems, the geometrical 
set hosting all the trajectories of a system is defined as the attractor 
\citep{eckmann1985ergodic}. Knowledge of the attractor informs on how often 
and for how long the trajectories of the system visit each region of the phase 
space and for how long the trajectory stays in the neighbourhood of each 
point. The 
behaviour of the system is entirely known if one can define those properties 
for each point of the attractor. In our set up, we do know the entire attractor of the turbulent atmosphere, but we can reconstruct important 
information on the dynamics via a projection on the two invariants $x_B$ and $y_B$. This projection is a subset of 
the attractor of the full turbulent system. It is a special Poincar\'e section because of the physical 
importance of the observables used to define it. In a similar fashion as in \cite{faranda2017stochastic}, we will 
study the dynamical properties of this Poincar\'e section and try to infer 
physical information on turbulence.\par
Our dynamical observables therefore consist of piece wise continuous 
trajectories of the turbulence anisotropy states defined by $(x_{B}(t) , 
y_{B}(t) )$, and the Poincar\'e 
recurrence theorem enables the analysis of properties of the attracting 
dynamics based on time series. The Poincar\'e recurrence theorem essentially 
states that certain dynamical systems, such as those bound to a finite volume, 
will after some time return to a state very close to the initial state. The 
time to return to an initial state depends on its location in phase space, and 
naturally on the required degree of closeness. A point $(x_{B}(t) , y_{B}(t) )$ 
in a timeseries of the invariants of the anisotropy tensor corresponds to a 
point in the attractor (the barycentric map or some part of it), and states 
whose distance to $(x_{B}(t) , y_{B}(t) )$ is small are the neighbours of that 
point or state. The density of points around each state $\zeta$, locally in space and 
time, defines a local dimension $d(\zeta)$ of the dynamics. In the barycentric map, the 
phase-space is the plane defined by the Poincar\'e section $(x_{B}, y_{B})$ 
and is thus 
two-dimensional. However if some states of anisotropy are visited less than 
others, locally the dimension $d$ of the Poincar\'e section may be smaller than 
two. For each anisotropy state $\zeta$, a local dimension $d(\zeta)$ can be quantified based on the timeseries $(x_{B}(t) , y_{B}(t) )$. 
Furthermore, if a trajectory leaves the neighbourhood of an initial state of anisotropy very 
fast, the persistence of the state will be small. If on the contrary the 
trajectory remains in the neighbourhood of the initial state for some time, the 
anisotropy state is more persistent. The persistence of an anisotropy state $\zeta$ is measured by an indicator $\theta(\zeta)$ defined as the inverse of the average persistence time in the neighbourhood of the anisotropy state $\zeta$. The indicator takes values $0 < \theta < 1$, where low values correspond to high persistence 
of the trajectory in the neighbourhood of $\zeta$, while values close to 1 
imply that the trajectory immediately leaves the neighbourhood of the anisotropy state $\zeta$. The methodology used to compute the values of the local dimension $d(\zeta)$ and the persistence $\theta(\zeta)$ is presented in the appendix and follows from \cite{Lucarini:2016ug}. \par
\section{Results and discussion}
\subsection{Scale interaction properties in classified flow regimes} \label{Sec:VARX}
We use the FEM-BV-VARX framework to classify flow regimes in the FLOSSII 
turbulence data. The turbulence data under consideration in Eq. (\ref{Eq:VARX}) 
is the vertical velocity variance $\sigma_{w}$ (defined in Section 
\ref{Sec:filter}) and the external factor is the sub-mesoscale wind velocity 
$V_{smeso}$ (Eq. \ref{Eq:Vsmeso}). The clustering analysis is performed based 
on the data collected from the height of two meters. This choice ensures that 
the data are within the boundary layer, which can be very shallow in strongly 
stable conditions. Investigations of the height dependence of flow regimes is 
left for future work. Here instead, the scale interaction properties will be 
analysed at different heights assuming that the regime affiliation is the same 
for all heights.\par 
User defined parameters of the framework include the maximum memory depth $p$ 
for the forcing variable $V_{smeso}$, the number of possible distinct VARX 
models or cluster states $K$ and the persistence parameter $C_{p}$, which 
limits the number of transitions between the states. The memory depth defines 
how many past states of the external factor $V_{smeso}$ are used in the model 
in Eq. (\ref{Eq:VARX}). The maximum memory depth that we use in this model is 
determined by a priori calculation of the partial autocorrelation function 
(\emph{pacf}) for the variable $V_{smeso}$ \citep{Brockwell:2002ur}. The 
correlation between the time series drops on average after a few minutes, and 
the memory depth is therefore set to $p=6$ (based on the average \emph{pacf} over 68 nights). To determine 
the optimum number of $K$ and $C_{p}$, multiple models are fitted for varied 
values of the parameters $K$ and $C_{p}$.\par 
In the clustering analysis of \cite{Vercauteren:2015fq}, the optimal model 
parameters were chosen as the minimisers of the Akaike Information Criterion (AIC). 
However for the FLOSSII dataset, the AIC exhibits asymptotic behaviour towards 
zero for all models in the investigated parameter space ($K=2,3,4,5,7$ and 
$C_{p}=[2,302]$) and cannot be used as a selection criteria. Instead, the 
optimal model parameters are selected as those that minimise the correlation 
between the signal $\sigma_{w}$ and the model residuals $\epsilon_{t}$, while 
maximising the amount of variance of the signal explained by the model. By 
observing the change of these two quantities over the parameter space, we found 
that increasing the parameters beyond $K=3$ and $C_{p}=150$ did not reduce the 
correlation in the residuals and did not increase the modelled variance. Thus 
the choice of $K=3$ and $C_{p}=150$ is considered as an optimal model. The 
amount of variance of $\sigma_{w}(t)$ explained by the VARX model in the three 
clusters is 0.8\%, 3\% and 9.5\%.\par
However, analysis of the model residuals showed that the error distribution in 
the cluster corresponding to the largest explained variance was not Gaussian. 
This cluster has the most interaction between sub-mesoscales and vertical 
velocity fluctuations as shown by the larger explained variance and we want to 
classify the dynamical interactions more accurately. Therefore, we select the 
time series in this specific cluster and classify it with the FEM-BV-VARX 
methodology further into two distinct clusters. This strategy leads to error 
distributions that are closer to normally distributed in the two subsequent 
clusters. The reason why this two-step procedure is helpful to the regime 
classification can be understood in the following way. The clustering procedure 
is based on minimising the euclidean distance between the data and the 
statistical model, under the constraint that the number of transitions between 
cluster states are bounded \citep{Horenko:2010gu}. Since a large part of the 
FLOSSII data show little dynamical interactions between $\sigma_{w}$ and 
$V_{smeso}$ (with 0.8\% resp. 3\% explained variance), the mean part of the 
statistical model ($\mu (t)$ in (\ref{Eq:VARX})) has the strongest effect in 
the overall distance minimisation. Indeed, inspection of the data classified in 
three clusters show that those correspond in large part to different mean 
values of $\sigma_{w}$. The periods of largest interactions between 
$\sigma_{w}$ and $V_{smeso}$ also correspond to very stable flow regimes 
\citep{Vercauteren:2015fq} with the smallest mean values of $\sigma_{w}$ and 
therefore these have the least weight in the distance minimisation. Selecting 
only those periods of larger dynamical interactions between $\sigma_{w}$ and 
$V_{smeso}$ enables a second level clustering which differentiates the 
dynamical interactions and not just the mean turbulent state. The fitted 
statistical models resulting from the two-step clustering strategy have a high 
degree of reproducibility. Over five repeated minimisation procedures for the 
FLOSSII data, the cluster affiliation function is consistent (or equal) to a 
degree of $90 \%$.\par
\begin{figure}[hbt]
	\centering
	\includegraphics{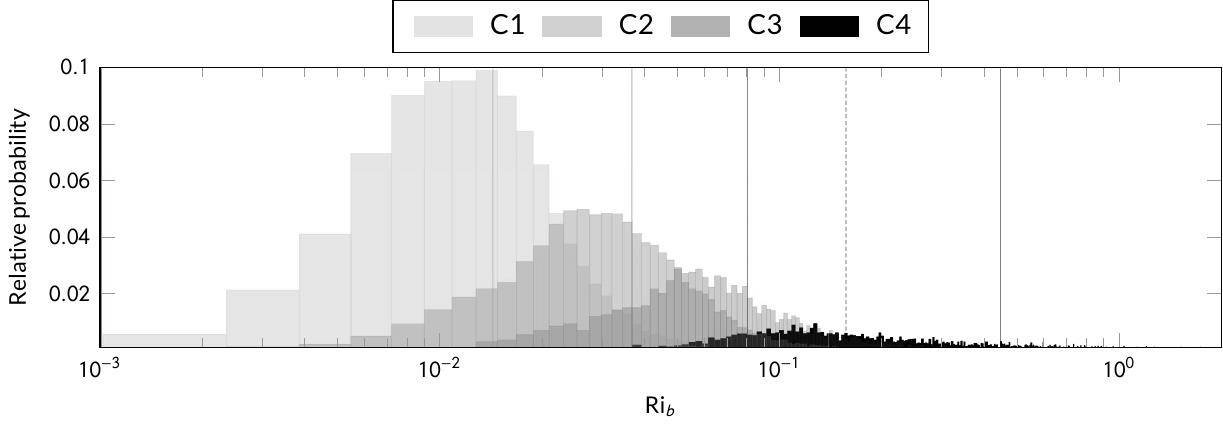}
	\caption{Histogram of the bulk Richardson numbers in the clustered flow 
	regimes. Vertical lines representing the median of the underlying 
	distribution, with the dotted line representing Ri$_{b(\text{crit})}$=0.25 for reference. Going from left to right the values are: 
	Ri$_{b(\text{C1})}$=0.03, Ri$_{b(\text{C2})}$=0.07, 
	Ri$_{b(\text{C3})}$=0.14, Ri$_{b(\text{crit})}$=0.25, 
	Ri$_{b(\text{C4})}$=0.59. For better illustration of the distribution 
	overlap the x-axis is showing a maximum Ri$_{b}=2$. The heavy tail of the 
	C4 distribution is reaching up to Ri$_{b(\text{C1})} = 6$ and explaining 
	the shifted value of the median.}   
	\label{fig:Rib} 
\end{figure}

 
As an indicator of the stability of the flow, we analyse the distributions of 
the bulk Richardson number
\begin{equation} \label{eq.: Rib}
\text{Ri}_b = (g/ \Theta_{0}) \frac{\left(T(z_{2}) - T(z_{1})\right) \Delta z }{\left(V(z_{2}) - V(z_{1})\right)^2 },
\end{equation}
in each cluster. $\Theta_{0}$ is the potential temperature averaged over 
all sensors and over the time of record (1 minute), $V$ is the record-averaged 
wind speed, $T$ is the record-averaged potential temperature derived from the 
sonic anemometer measurements,  
$\Delta z$ is the difference in height between the two levels $z_{1} = 1\, \text{m}$ and $z_{2}= 10\, \text{m}$ and $g$ is the 
gravitational acceleration. The $\text{Ri}_b$ distributions conditional on the 
four identified flow regimes are shown in Fig. \ref{fig:Rib}. The bins of the 
distributions are normalised with the number of samples in each regime to 
obtain the relative probability. The clustering strategy is found to separate 
the $\text{Ri}_b$ distributions into values indicative of weakly stable flows 
and strongly stable flows, albeit with large overlaps in the distributions. The 
distributions of C1, C2 and C3 are located well below the 
Ri$_{b(\text{crit})}$=0.25 and the boundary layer state at that times can be 
distinctly interpreted as weakly stable. The strongly stable cluster C4 shows a pronounced heavy tail decaying towards Ri$_{b}$=6 (not shown) and is showing a significant spread around the Ri$_{b(\text{crit})}$. When going from weakly stable cluster towards strongly stable clusters, the distribution of the Ri$_{b}$ tends to diffuse, partly due to low values of the shear velocity that lead to uncertainties. The overlap of distributions and the diffused distribution of Ri$_{b}$ highlights the difficulty of defining a threshold based on Ri$_{b}$ for distinguishing flow regimes. As a note of caution, the distributions in Fig. \ref{fig:Rib} should be considered only as a qualitative indication of stability properties, since the temperature from the sonic anemometers are known to experience drift and biases between sensors exist, leading to uncertainties in the values.  Hence we discard presenting a detailed analysis of the Ri distributions at each measurement height. Still, Fig. \ref{fig:Rib} shows that the clustering strategy, similarly to what was found in \cite{Vercauteren:2015fq}, separates periods of qualitatively different stability according to Ri$_{b}$.
\par
\begin{figure}[hbt]
	\centering
	\includegraphics{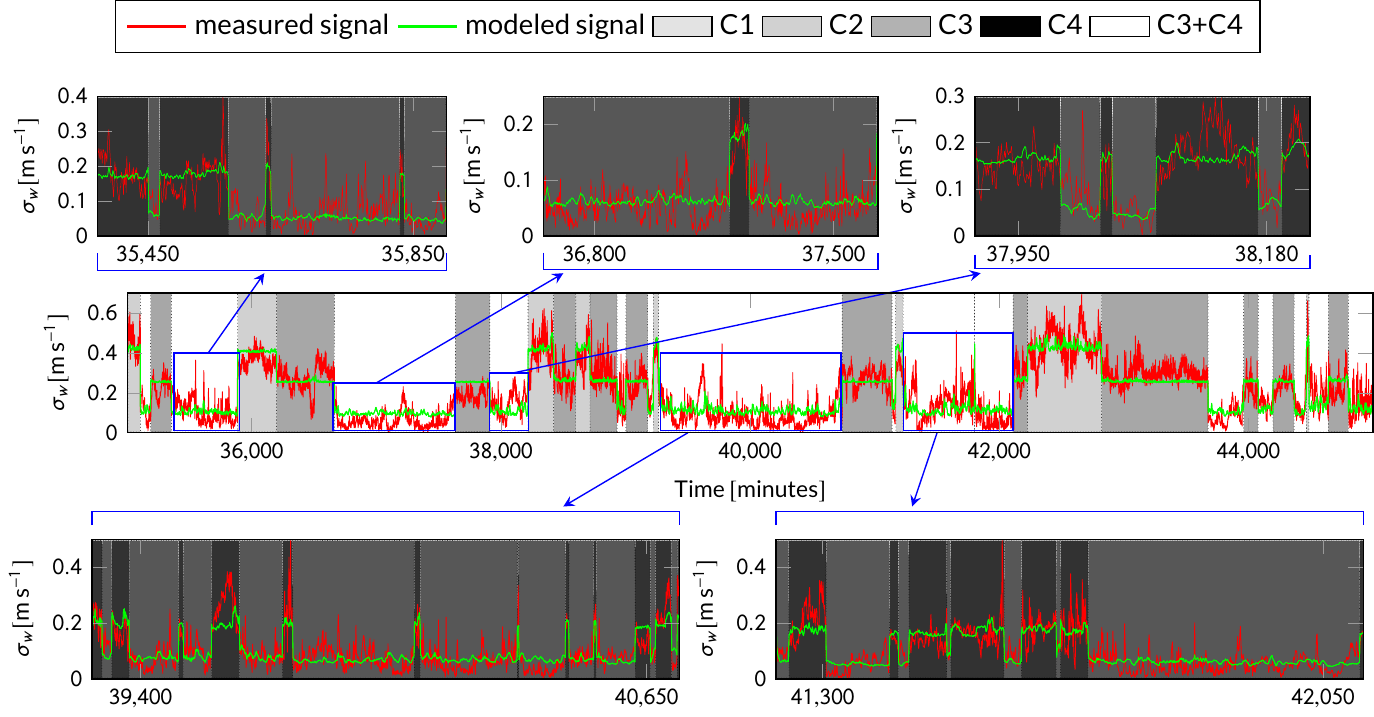}
	\caption{Example of measured and modelled timeseries and flow regimes. The middle horizontal 
	panel represents a period of 68 nights. The inserts illustrate the solution produced by  
the second application of the clustering methodology.}
	\label{FIG:timeseries}
\end{figure}

 The clustering of the time series of $\sigma_{w}$ using $V_{smeso}$ as 
 external factor is shown for an example period in Fig. \ref{FIG:timeseries}, 
 where the background colours denote regime affiliation. The middle horizontal 
 panel show the first clustering procedure with three clusters. The Cluster 
 (C3+C4) with the lowest mean is then considered as one continuous time series  
 and clustered again to result in C3 and C4. The solution of this second 
 clustering procedure is then illustrated in the inserts panels. The cluster 
 (C3+C4) is not considered for the following analysis and is shown here for 
 explanatory reasons. By comparing the modelled time series between the middle 
 panel and the inserts panels, one notices that for the strongly stable 
 condition (namely comparing C3+C4 vs C3 and C4) the sub-clustered solution is 
 describing the mean  better compared to the C3+C4 solution. To achieve the same performance with one level clustering we needed at least seven clusters. The solution 
 with that number of clusters started to be unreproducible, meaning that the 
 affiliation functions diverged for different solutions.\par 
 In Fig. \ref{FIG:timeseries} the dynamics of $\sigma_{w}$ is poorly 
 captured by 
 the model in 
 the periods with more mixing (regimes C1 and C2). As a more 
 quantitative indication of the differences in scale interactions, we compare the VARX model coefficients estimated for regimes C1 to C4. The VARX model in Eq. (\ref{Eq:VARX}) contains a total of 9 parameters: $\mu$ corresponds to the mean value of $\sigma_w$, $B_0$ to $B_6$ are the weights associated with the past history of the external factor $V_{smeso}$ and $C$ is the weight associated to the noise part of the model. In order to compare the relative weight of the mean versus the external factor in each statistical model, we normalise each parameter by the largest one. In all models, the largest parameter is $\mu$. We then compute the norm $B_{Ci}$ of the vector $\left( B_0/\mu, \cdots, B_6/\mu \right)$, where $B_{Ci}$ is associated to the model coefficients in cluster C$i$, to estimate the relative weight of the external forcing in each statistical model. Note that with this normalisation, the weight of the mean is always 1. The values obtained for C1 to C4 are, in order, $B_{C1}=0.04$, $B_{C2}=0.04$, $B_{C3}=0.11$ and $B_{C4}=0.29$. The increasing values denote that the more stable cases show more statistical causality between non-turbulent scales of motion and turbulence. Although the statistical model for $\sigma_w$ is arguably insufficient, the method captures subtle differences in scale interactions between different regimes. Indeed, our goal here is not to represent $\sigma_w$ accurately, but rather to capture transitions between different states of the SBL and the method appears appropriate for that goal. \par 
\par
\begin{figure}[hbt]
	\centering
	\includegraphics{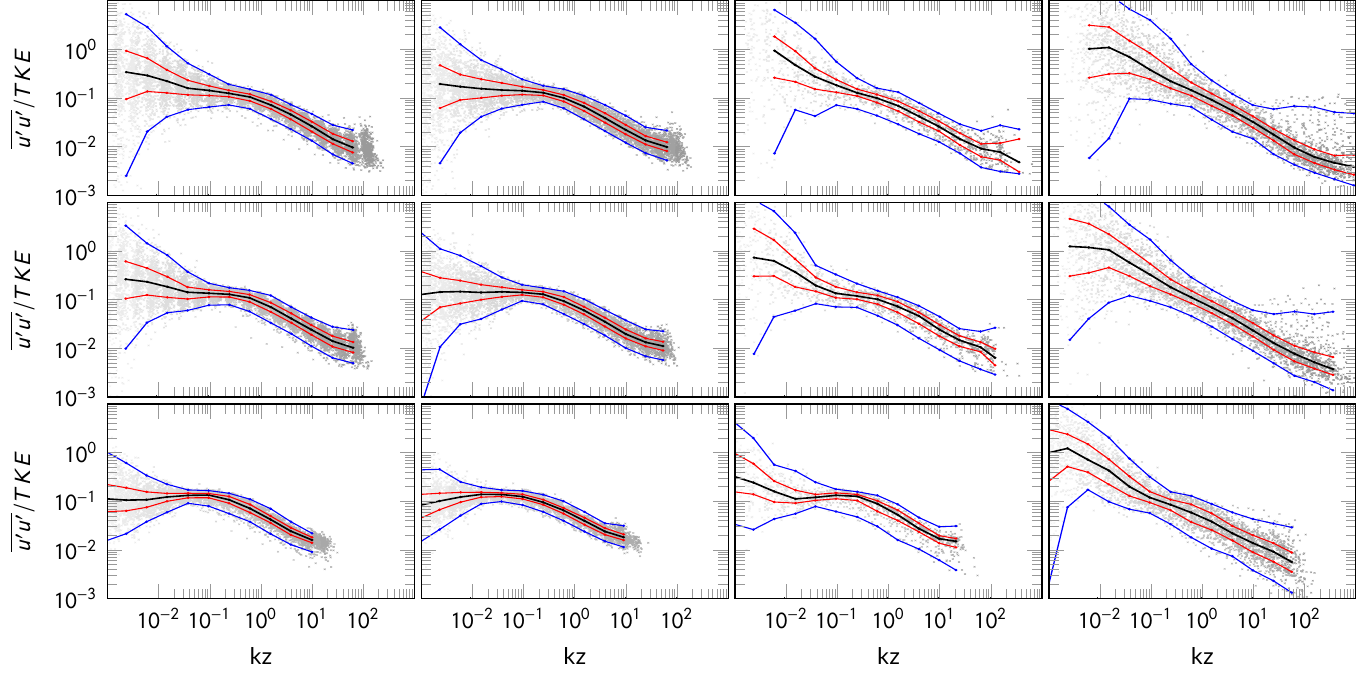}
	\caption{MRD energy spectra of the streamwise velocity component. The top, middle and bottom rows correspond respectively to the measurement heights of 30 meters, 15 meters and two meter. Observing from left to right, every column of panel corresponds to regimes C1-C4 respectively. C1 is a weakly stable regime 
and C4 is a strongly stable regime. Wavenumbers $k$ are calculated assuming Taylor's frozen turbulence hypothesis and $z$ is the measurement height. The energy is normalised by the average TKE calculated based on one-minute bins in order to consider turbulent scales only. The median spectra are represented by the black line, the red lines
represent the 25th and 75th percentiles, and the blue lines show the 5th and 95th percentiles. Grey dots represent individual values.} 
	\label{fig:MRD_u}       % Give a unique label
\end{figure}
\begin{figure}[hbt]
	\centering
	\includegraphics{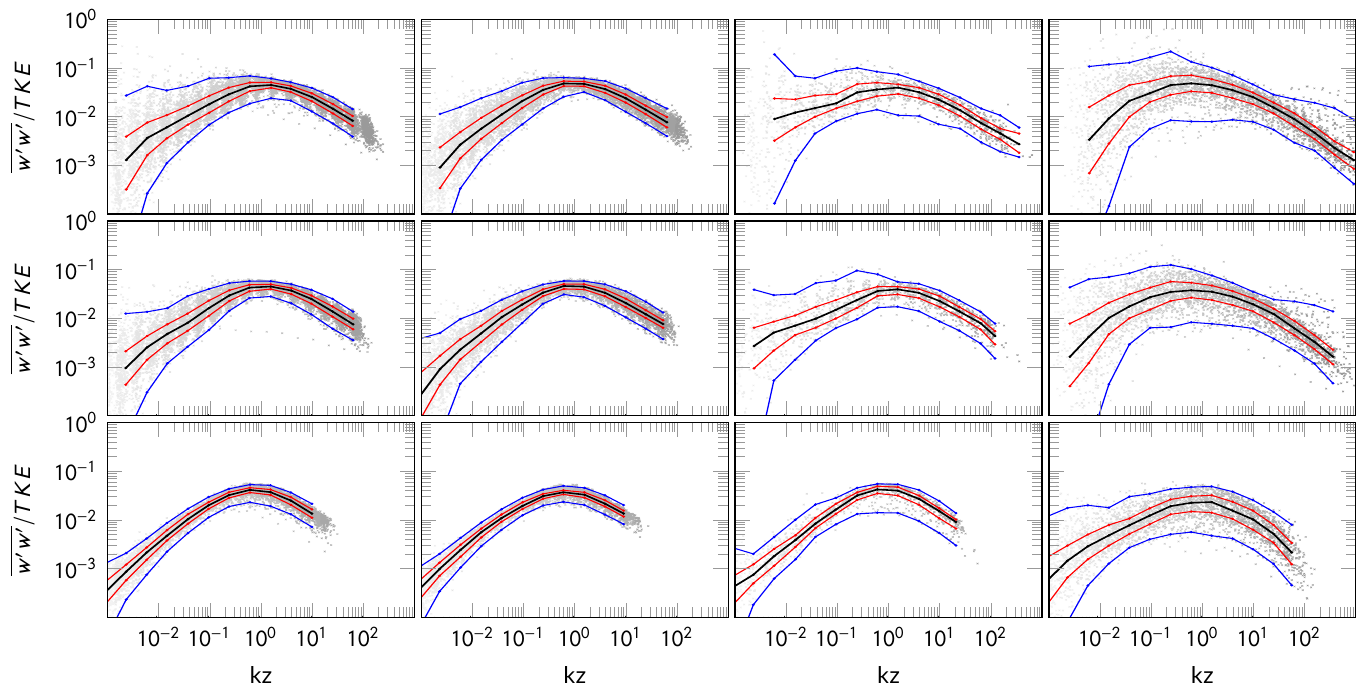}
	\caption{MRD energy spectra of the vertical velocity component. The top, middle and bottom rows correspond respectively to the measurement heights of 30 meters, 15 meters and two meter. Observing from left to right, every column of panel corresponds to regimes C1-C4 respectively. C1 is a weakly stable regime 
and C4 is a strongly stable regime. Wavenumbers $k$ are calculated assuming Taylor's frozen turbulence hypothesis and $z$ is the measurement height. The energy is normalised by the average TKE calculated based on one-minute bins in order to consider turbulent scales only. The median spectra are represented by the black line, the red lines
represent the 25th and 75th percentiles, and the blue lines show the 5th and 95th percentiles. Grey dots represent individual values. }
	\label{fig:MRD_w}       % Give a unique label
\end{figure}

To illustrate how the total energy is distributed among 
the scales of motion under different near-surface SBL conditions, 
multi-resolution decomposition (MRD) energy spectra are shown for the heights 
of 2 m, 15 m and 30 m in Fig. \ref{fig:MRD_u} (for the streamwise velocity component) and Fig. 
\ref{fig:MRD_w} (for the vertical velocity component) for the four classified flow regimes C1-C4 
(from left to right). The wavenumbers are calculated assuming Taylor's frozen turbulence hypothesis, and the energy spectra are scaled by the TKE averaged based on bins of one minute. In this way, the normalisation considers only the energy content of the turbulent scales, discarding the sub-mesoscales. From Fig \ref{fig:MRD_u}, it is visible that the energy content of the streamwise velocity component on sub-mesoscales is highly variable in all flow regimes. The spread among cases increases with height in weakly stable flow regimes (C1 and C2), but is large for all heights in very stable regimes (this is mostly the case in C4). The median exhibits a pronounced plateau for scales larger than $kz = 1$ in regimes C1 and C2. In C3 such a plateau around $kz =1$ is still present, albeit significantly smaller. In C4, on the other hand, the increase of energy with increasing scales is continuous, denoting the absence of a spectral energy gap between the turbulence and the sub-mesoscales. This small or absent scale gap between submeso motions and turbulence in the very stable regimes C3 and C4 is a sign of unsteady forcing of turbulence by sub-mesoscales with very variable energy content. We can speculate that this is a cause for the observed larger scatter in the scales smaller than $kz = 1$ in C3 and C4. In the energy spectra of the vertical velocity component (Fig \ref{fig:MRD_w}), more scatter is apparent at all scales in C3 and C4. Likewise, this may be explained by the unsteady forcing that hinders the formation of a universal inertial subrange. The variability of the energy content of the sub-mesoscales also increases with height, but is much smaller than that of the horizontal component.

\subsection{Anisotropy characteristics in different flow regimes}
Having characterised the scale-wise energy content of the velocity components in classified flow regimes, we can now turn to the analysis of anisotropy characteristics of the Reynolds stresses. In order to calculate the anisotropy tensor (Eq. 
\ref{Eq:anis}) and its invariants, an averaging scale has to be defined. From 
observations of MRD heat flux cospectra (not shown) it appears that the heat flux cospectra 
level-off at averaging times ranging between approximately 5 minutes (C1) and 1 
minute (C4), depending on how strongly stable the flow regime is. In order to 
minimise contributions from sub-mesoscale motions in the anisotropy analysis, 
we select the shortest averaging timescale of 1 minute as was done in 
\cite{Stiperski:2017db}. Note that this choice implies that the anisotropy 
analysis will discard some of the turbulent contributions to the anisotropy tensor 
in the less stable flow regimes.\par
\begin{figure}[tbp]
	\centering
	\subfigure[Cluster 1]{\includegraphics{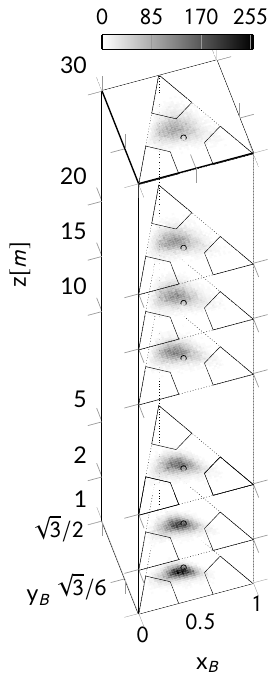}}
	\subfigure[Cluster 2]{\includegraphics{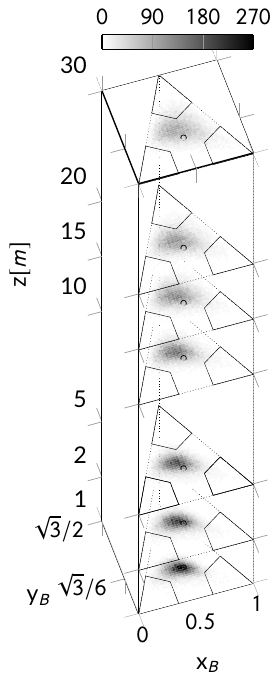}}
	\subfigure[Cluster 3]{\includegraphics{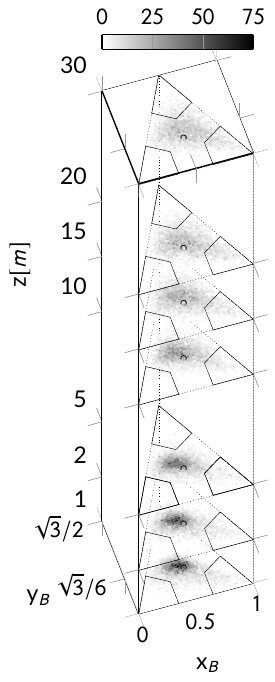}}
	\subfigure[Cluster 4]{\includegraphics{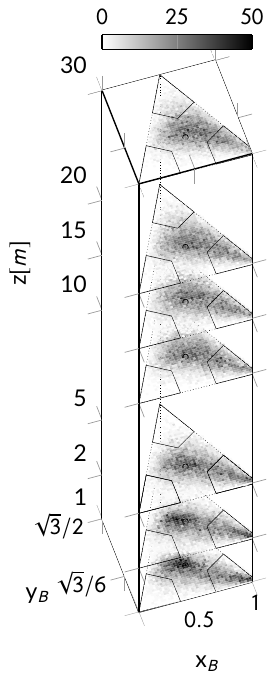}}	
	\caption{Anisotropy as states in the barycentric map for each height and 
	flow 
regime. The colourbar shows the density of points in each state of the 
barycentric map.}
	\label{fig:anisotropy towers}       % Give a unique label
\end{figure}

The distribution of anisotropy states is shown for each cluster C1-C4 and each 
measurement height in Fig. \ref{fig:anisotropy towers}, where the grey scale 
shows the density of points. While mixed states of anisotropy, i.e. towards the 
middle of the barycentric map, are the most common in all cases, marked 
differences appear in the limiting states. Here we follow 
\cite{Stiperski:2017db} to define limiting "pure" states of anisotropy as 
states falling in edges of the barycentric map, where the limiting lines for 
each edge were chosen to cover 70 \% of the sides of the equilateral triangle 
as illustrated in Fig. \ref{Fig:BAM}. The results, however, do not show large sensitivity to this choice. The isotropic states correspond to the 
upper corner of the barycentric map. A height dependance is clearly apparent 
here. Isotropic stresses are only found away from the ground, so that the higher levels have 
the highest densities of isotropic stresses, regardless of flow regime 
affiliations. This result is not surprising since the presence of the ground 
surface enhances the shear distortion effects on turbulence and limits its isotropy, as was 
discussed elsewhere (e.g. \cite{Antonia:2001wb}). The shear distortion effects in the 
absence of thermal stratification typically lead to two-component 
stresses due to the straining effect. Accordingly, two-component stresses were observed to be prevalent near the wall during daytime, unstable conditions with active turbulence by \cite{Stiperski:2017db}. In low wind speed, very stable conditions however, the shear generation of turbulence may be too weak to sustain active turbulence coupled to the ground surface and such shear distortion effects may thus not be prevalent. Therefore, although all stresses appear closer to axisymmetric states when one approaches the ground surface in our dataset, clearly the preference is towards one-component stresses. 
This preference depends on regime affiliation. When analysing TKE budget terms during periods of one-component stresses, \cite{Stiperski:2017db} found that the source of turbulence was mainly non-local, likely due to advected turbulence. Our results are in agreement with this finding since one-component stresses are the most abundant when sub-mesoscales are most active, and submeso motions represent a means of turbulent transport and non-local generation. \par

\par
\begin{figure}[bt]
	\centering
	\includegraphics{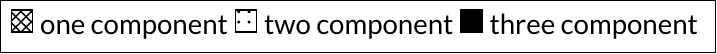}
	\includegraphics{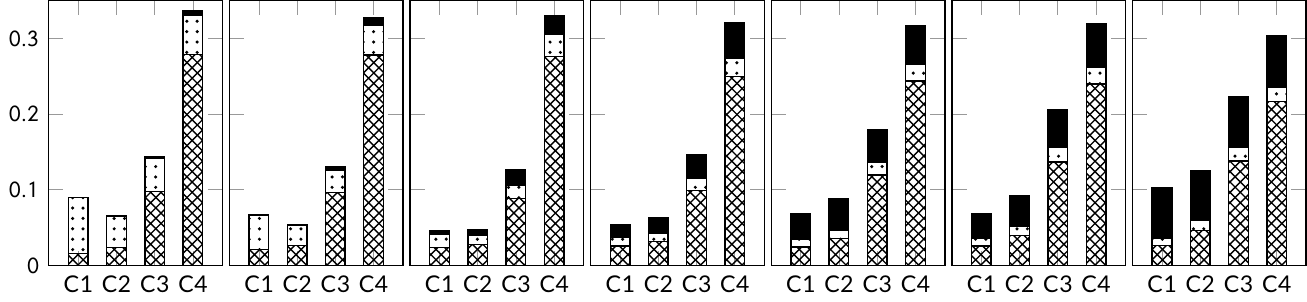}
	\includegraphics{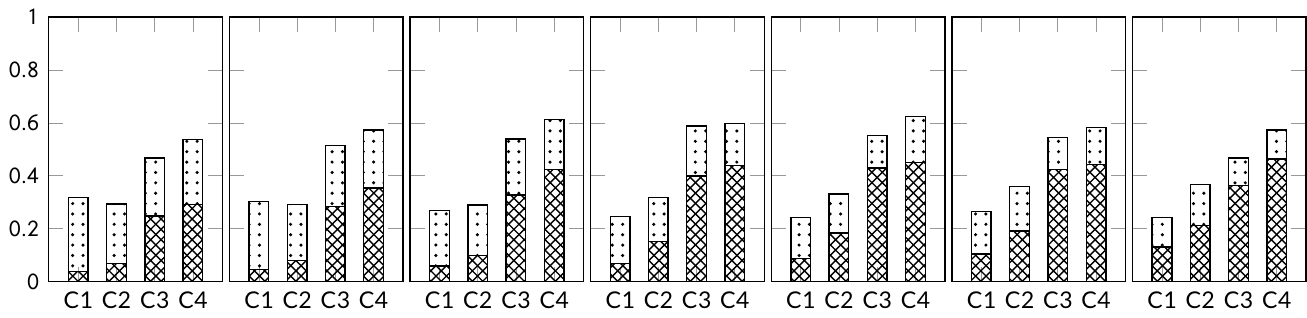}
	\caption{Occurrences of each anisotropy state for different heights and 
clusters. Top panels: one-minute averaged Reynolds stresses. Bottom panels: 30-minute averaged Reynolds stresses. Left to right: 1, 2, 5, 10, 15, 20 and 30 m above ground. The numbers 
1,2,3,4 denote the flow regime affiliations C1-C4. The length of the bar 
represents the percentage of stresses within each edge. The pure states are defined as in 
Fig. \ref{Fig:BAM}} %\ref{Fig:BAM}
	\label{fig:limitingstates}
\end{figure}

The density of stresses in the one-component limiting states becomes 
higher for increasing regime affiliation number (Fig. \ref{fig:anisotropy 
towers}) corresponding to increasingly stable conditions (Fig. \ref{fig:Rib}), 
and increasing influence of sub-mesoscale motions as discussed above. We 
specifically quantify the percentage of stresses falling in each limiting state 
of anisotropy in the top panels of Fig. \ref{fig:limitingstates}, conditional on regime 
affiliation and height of measurement. The isotropic edge (black bar in the 
figure) is almost absent until the measurement height of 5 m, and then the 
proportion of isotropic stresses increases with height for all flow regimes, 
having almost the same proportion in all regimes (with the maximum at 30 m of 
slightly less than 10 \%). The proportions of two-component stresses show an 
opposite trend, decreasing with height but being also very similar for all flow 
regimes. This is due to the small scales of turbulence considered here. By calculating the Reynolds stresses based on one-minute averages, shear distortion effects leading to two-component stresses are limited to correspondingly close distances to the wall. The proportions of one-component stresses also decreases for 
increasing 
heights, but here the regime dependence is strong. In C1 and C2, the  
proportions of one-component stresses  is very small at all heights, while it 
is 
large in C3 and even large in C4, reaching almost 30 \% of the states near the 
surface. Modifying the choice of threshold used to delineate the "pure" anisotropy states (excluding 20\% and 40\% of the barycentric map) leads to qualitatively similar results, with the proportions being all larger or smaller, respectively. 

In regimes C3 and C4, the energy content of the sub-mesoscales is often larger than the energy content of the turbulent scales (Fig. \ref{fig:MRD_u}) and is largest in the horizontal velocity component. Since the activity of these sub-mesoscales occurs on scales 
just above or similar to the largest turbulent scales, the turbulence is 
forced by unsteady submeso motions and therefore continuously deformed. The isotropisation is presumably hindered by the constantly changing anisotropic forcing. In order to estimate the anisotropy of the submeso motions, we compare the proportions of limiting anisotropy states obtained from 30 min averaged Reynolds stresses (Fig. \ref{fig:limitingstates}, bottom panels). At that scale, as expected eddies are never isotropic, however, the two-component limit is much more frequent, especially close to the surface. This is to be expected due to the presence of the wall that impacts the larger eddies considered at this scale. It can also be expected that the higher proportion of two-component turbulence is due to mixed states that are more axisymmetric at larger scales. The proportion of one-component limiting states is also larger than the corresponding proportions for the 1-minute averaged stresses, and increases for increasingly stable regime similarly to the 1-minute averaged stresses. Thus we might conclude that submeso motions can be both two-component axisymmetric and one-component, but the turbulence they produce can still be at times isotropic, as well as highly anisotropic. This leads us to hypothesise that the anisotropy of the forcing submeso motions transfers to anisotropic turbulence even at the scale of one minute considered here. This is likely due to the lack of scale separation between the turbulence and its unsteady anisotropic forcing, such that the turbulence cannot equilibrate to an isotropic state. \par
\par
In order to facilitate interpretation of the one-component stresses in the physical space, we calculate the 
non-dimensional velocity aspect ratio (VAR) for the limiting one-component stresses and for all other cases separately. This ratio takes the value of one if all three standard deviations approach the same value and is defined by \cite{Mahrt:2012do} as:   
\begin{equation}
	VAR \equiv \frac{\sqrt{2}\sigma_w}{\sqrt{\sigma_u^2 + \sigma_v^2}} \, .
\end{equation}
In Tab. \ref{tab.:VAR} we evaluate the mean value of the VAR in each regime for 
cases corresponding to the one-component anisotropy state and compare it to 
mean value of VAR for the periods that are not in the one-component anisotropy 
state. As we change the regime from C1 to C4 we observe a decrease of the VAR 
from 0.10 to 0.07, along with an increase of the standard deviation of VAR. In comparison, the 
ratio 
for cases outside the one-component limit is not dropping below 0.20. Thus the vertical component of the Reynolds 
stress is smallest for the one-component cases, and reduces for 
increasing regime affiliation number. 
\begin{table}[h]
\caption{Mean and standard deviation of VAR for one-component (top row) and non 
one-component (bottom row) anisotropy states for regimes C1-C4 at the height 
of two meters.}
\centering
\begin{tabular}{c|cccc}

component & C1 & C2 & C3 & C4 \\ 
\hline 
one & 0.10 $\pm$ 0.03 & 0.10 $\pm$ 0.03 & 0.09 $\pm$ 0.04 & 0.07 $\pm$ 0.05 \\ 

non one & 0.20 $\pm$ 0.05 & 0.23 $\pm$ 0.05 & 0.23 $\pm$ 0.07 & 0.20 $\pm$ 0.10 
\\ 

\end{tabular} 
\label{tab.:VAR}
\end{table}

\subsection{Anisotropy characteristics of counter gradient cases}
\begin{figure}[tbp]
	\centering

	% Figure left--------------------------------------------------
	\subfigure[With gradient. \label{fig:countergrad:sub:counterGrad}]
	{ 	
		\includegraphics{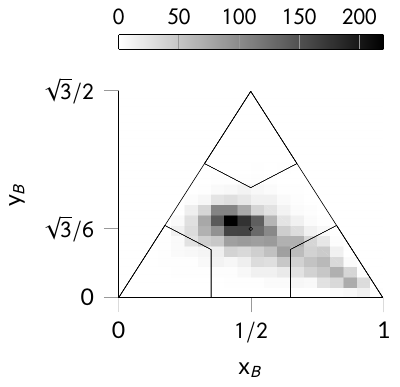}	
    }\quad
    %-------------------------------------------------------------
    %
	\subfigure[Heat Flux cospectrum.\label{fig:countergrad:sub:HF}]
	{
		\includegraphics{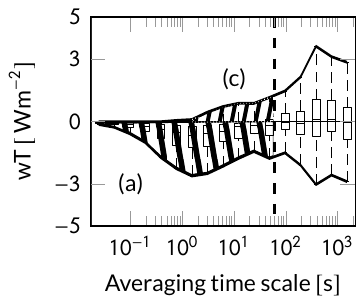}
	}\quad    
	%
	% Figure right -----------------------------------------------------    
	\subfigure[Counter gradient.\label{fig:countergrad:sub:withGrad}]
	{
		\includegraphics{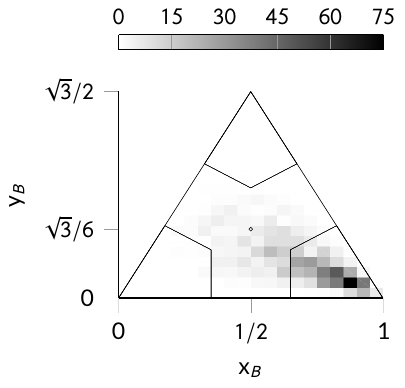}
	}\quad
	%-------------------------------------------------------------------
	%
	\caption{Anisotropy dependence on the sign of sensible heat flux (b) in C4 regime on height of two meters. In (a) and in (c) the color map is showing the density of points.  }    
	\label{fig:countergrad}       % Give a unique label
\end{figure}

Interactions between waves or sub-mesoscale motions and turbulence have been 
shown to lead to counter-gradient fluxes \citep{Einaudi:1993ut}. In Fig. 
\ref{fig:countergrad}, we separate the stresses falling in the flow regime C4 
into two categories, namely periods of negative sensible heat flux (on one 
minute averaging scale) and periods of positive (counter-gradient) sensible 
heat flux as shown in Fig. \ref{fig:countergrad:sub:HF}. The temperature gradient is always negative, i.e. stratification is stable for all considered cases. To separate the 
anisotropic state in Fig. \ref{fig:countergrad:sub:withGrad} and Fig. 
\ref{fig:countergrad:sub:counterGrad} we use the heat flux 
$\overline{w^{\prime}T^{\prime}}$ averaging scale of one minute.  The peak of 
the density of anisotropy states for the cases of negative sensible heat flux 
(with-gradient) occurs in the middle of the barycentric map. In the 
counter-gradient cases however, the peak of the distribution lies within the 
edge corresponding to the one-component limiting states. Hence, most of the 
counter-gradient cases correspond to one-component limiting states. In order to 
analyse if the reverse is true, i.e. if one-component limiting states are 
mainly counter-gradient cases, the percentage of cases with positive and 
negative sensible heat flux is listed in Tab. \ref{tab.:oneComVsGradient} for 
all one-component limiting states in each flow regime C1-C4. The values show 
that only a small proportion of cases in one-component limiting states 
correspond to positive sensible heat flux, and that the percentage is similar for all flow regimes, so that it is not possible to associate one-component turbulence with a specific set of submeso motions which would cause counter gradient fluxes. 
 
\begin{table}[tbp]
\centering
\caption{ Percentage of cases of negative (+g for with gradient) and positive (-g for counter gradient) sensible heat flux (one minute scale) observed in the one component limiting anisotropy state (right corner of the barycentric map). For the column 'total' no affiliation function is involved. The affiliation function to determine regimes C1-C4 is calculated based on height of two meters and is used to evaluate the table entries for all heights.   }
\begin{tabular}{c|ccccc}
~ & C1 & C2 & C3 & C4 & total \\  
height [m] & +g [$\%$] / -g [$\%$]  & +g [$\%$] / -g [$\%$] & +g [$\%$] / -g [$\%$] & +g [$\%$] / -g [$\%$] & +g [$\%$] / -g [$\%$] \\ 
\hline 
30 & 84.56/15.44 & 86.17/13.83 & 83.87/16.13 & 86.67/13.13 & 85.88/14.12 \\ 

15 & 90.34/9.66 & 90.79/9.21 & 89.15/10.85 & 91.53/8.47 & 90.95/9.05 \\ 

2 & 97.24/2.76 & 93.96/6.04 & 97.69/2.31 & 96.20/3.80 & 96.17/3.82 \\ 

\end{tabular} 
\label{tab.:oneComVsGradient}
\end{table}

\subsection{Dynamical indicators in the anisotropy dynamics}
\begin{figure}[tbp]
	\centering
	\subfigure[C1]{\includegraphics{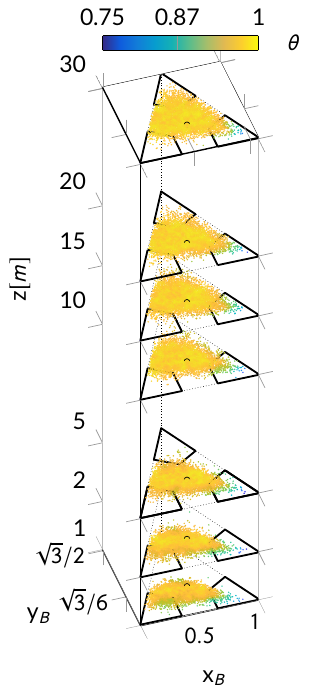}}
	\subfigure[C2]{\includegraphics{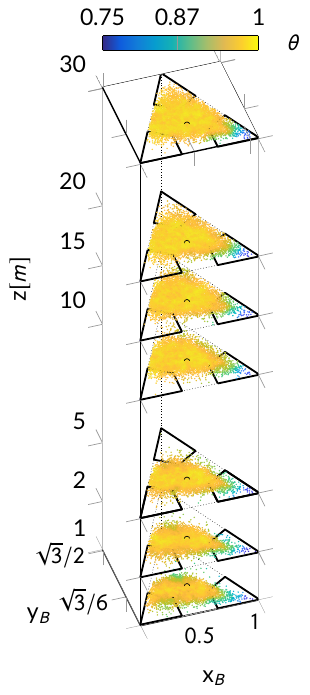}}
	\subfigure[C3]{\includegraphics{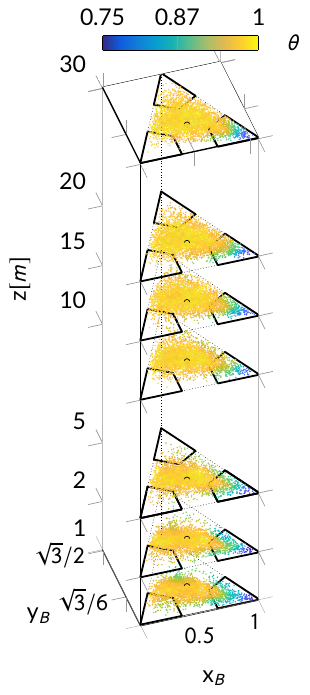}}
	\subfigure[C4]{\includegraphics{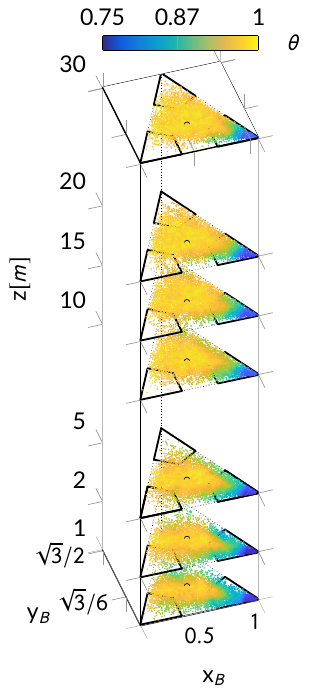}}
	\caption{Persistence of the dynamics of the anisotropy state. The colourbar 
	is indicating the value of $\theta$ , where low values correspond to high 
	persistence of the trajectory in 
	the neighbourhood and values close to 1 imply that the 
	trajectory immediately leaves the neighbourhood. }
	\label{Fig:persistency}
\end{figure}

We now turn to the analysis of the dynamics of the states of anisotropy. We 
want to investigate if the rate of isotropisation depends on the initial 
anisotropy state, and on the background flow regime. Moreover we are interested 
in assessing the trajectories of the stresses in the anisotropy invariant 
coordinates. Based on the timeseries of the anisotropy invariant coordinates 
$x_{B}$ and $y_{B}$, we estimate the persistence and dimension of the dynamics 
as presented in section \ref{Sec:persistency}. Figure \ref{Fig:persistency} 
shows a scatterplot of the persistence indicator $\theta$ estimated from the 
parameter of the distribution in Eq. \ref{Eq:probExt}, the colour showing the 
value of the indicator. It is obvious that $\theta$ values are smaller in the 
edge corresponding to one-component stresses, denoting longer-lived states. We 
recall that values close to $\theta=1$ imply that the trajectory immediately 
leaves the initial anisotropy state, while smaller values denote that the 
dynamics resides in a neighbourhood of the initial state. In the mixed states 
in the centre of the barycentric maps, the indicator values are very close to 1 
denoting that those states are modified almost instantaneously. 
The fact that one-component stresses are long-lived is in accordance with the \cite{KWINGSOCHOI:2018hj} finding for homogeneous turbulence, if one takes one-component turbulence to be an asymptotic limit of cigar shaped turbulence.\par

\begin{figure}[tbp] 
	\centering
	\subfigure[C1]{\includegraphics{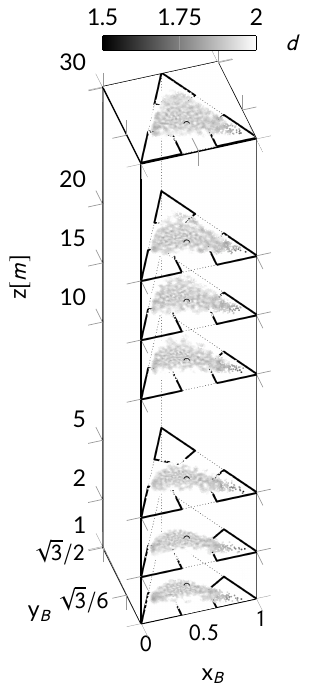}}
	\subfigure[C2]{\includegraphics{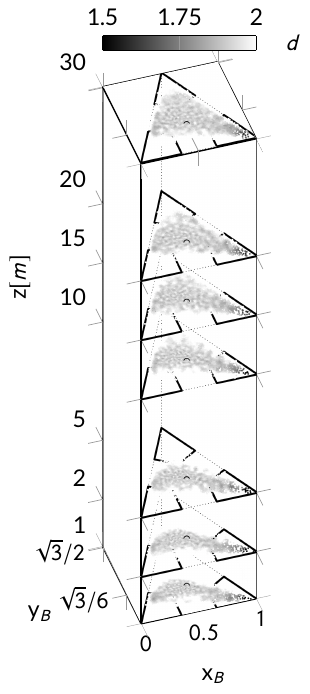}}
	\subfigure[C3]{\includegraphics{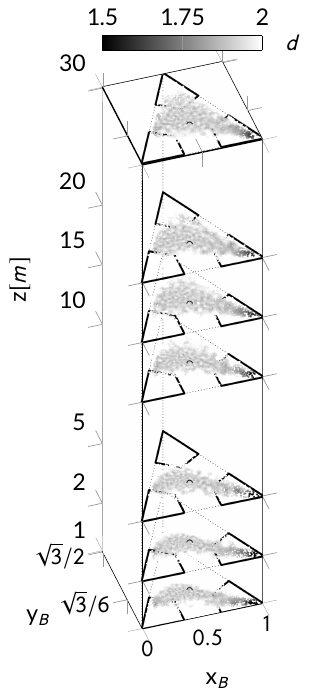}}
	\subfigure[C4]{\includegraphics{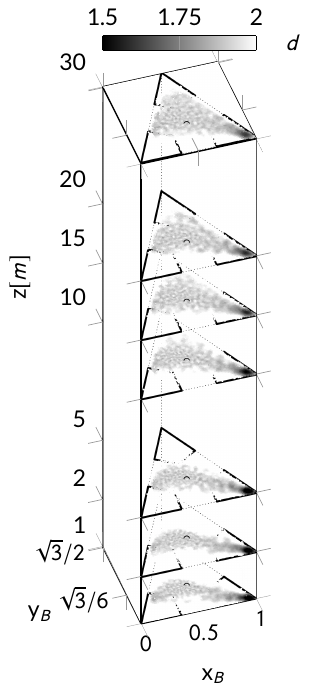}}
	\caption{Local dimension of the anisotropy states.}
	\label{Fig:dimension}
\end{figure}

We calculate the estimates of the local dimension of the dynamics based on Eq. 
\ref{Eq:dim}, for each stress and show the scatterplot of the estimated 
dimensions in Fig. \ref{Fig:dimension}. The maximum local dimension is two, 
since the phase-space is the plane formed by the eigenvectors, i.e. $x_{B}$ and 
$y_{B}$. A local dimension of two indicates that there is no preferred 
direction in which the anisotropy state is altered; it can change in any 
direction on the plane, starting in an initial state. On the contrary, a 
smaller 
local dimension indicates that the way that the anisotropy state is modified 
occurs in a restricted part of the plane, that is, with a preferred direction. 
This is evidently the case close to the one-component edge of the barycentric 
map, and along a line connecting the one-component edge to the centre of the 
map. This denotes a preferential path away or towards one-component 
anisotropy. It is interesting to note that this preferential path does not 
include or evolve towards two-component axisymmetric states, but shows that intermittent bursts of 
turbulence are mostly of 
axisymmetric oblates, with a more pronounced third direction. \par
The combination of persistence and dimension analysis possibly shows that 
one-component stresses are more stable topologically, and that the formation or 
destruction of such topological structures takes a preferential route. 
\par
\section{Conclusions}
We classified SBL flow regimes based on the intensity of turbulent velocity fluctuations and their modulation by a sub-mesoscale wind velocity. The anisotropy properties of the Reynolds stress tensor were then investigated in each thus classified flow regime. This combination of methods highlighted different properties of the turbulent stresses under the influence of submeso motions. We showed that the influence of submeso motions on turbulence gains significance as the stability increases, and that the more submeso-influenced flow regimes have a marked preference for one-component axisymmetric stresses. This topological signature is in part resulting from the buoyancy damping effects as the Richardson number increases beyond the critical value (cf. Fig 2), but also from the influence of sub-mesoscale forcing of turbulence. Close to the surface (below 2 m), the effect of shear competes with buoyancy damping effect, resulting in a larger proportion of two-component axisymmetric stresses, a typical signature of shear induced smaller-scale turbulence, but also the effect of the wall on the larger 30 min sub-mesoscale eddies. Small-scale isotropic stresses were found at higher levels in all flow regimes, where shear effects and the wall are no longer felt by turbulent eddies. 

The present results are unable to completely elucidate the differences between the one-component and two-component turbulence found by \cite{Stiperski:2017db}, but show that the proportion of one-component cases increases for increasing energy content of the sub-mesoscale dynamics. Furthermore, the regimes with most energy content in the sub-mesoscales are characterised by a lack of scale separation between the turbulence and the sub-mesoscales. We hypothesised that this lack of scale separation is the cause for the prevalence of small-scale one-component turbulent stresses that are driven by anisotropic submeso motions. Confirming this hypothesis, we found that a large proportion of the periods where the sensible heat flux was against the mean downward gradient, typically corresponding to submeso motions, was characterised by one-component limiting states of anisotropy. Yet only a small fraction of all one-component limiting cases happen during counter-gradient periods. Submeso motions can take a variety of form that may not lead to counter gradient fluxes but still impact the topology of the turbulence. The existence of almost isotropic turbulence even in the most sub-mesoscale influenced regime shows the unsteady nature of this regime, and that even one-component or two-component axisymmetric eddies at 30 minute scale can initiate bursts of small-scale isotropic turbulence, as previously observed by \cite{Stiperski:2017db}. 
\par

Additionally we showed that one-component stresses were more persistent in  their dynamics, and highlighted signs of a preferred route towards or away from one-component stresses in the topological state space. This route interestingly does not involve purely two-component axisymmetric turbulence, but is more of a axisymmetric oblate type. However, \cite{Stiperski:2018JD} show that the lack of small scale two-component axisymmetric turbulence is often observed in other datasets. An interesting future analysis would be to investigate the scale-wise return to isotropy in parallel with our results on the persistence and dimension of the dynamics.

The results can be used to improve the representation of non-stationary turbulence under the influence of sub-mesoscale motions. This is pertinent for subgrid-scale turbulent parameterisation, which are currently mainly based on isotropic eddy diffusivity models. Our results show that anisotropic modelling is required in cases where the variability of sub-mesoscales is important in relation to the turbulent scales. The signs for a preferred route towards or away from the one-component stresses despite strong influence of random sub-mesoscale motions provide encouraging perspectives for representing the return-to-isotropy in future models. \par 
\section*{acknowledgements}
The authors would like to thank Marc Calaf for inspiring discussions at an early stage of this analysis. 
Larry Mahrt is acknowledged for providing the FLOSSII data and his help to 
analyse them. Illia Horenko and Dimitri Igdalov provided the FEM-BV-VARX
framework and help that was greatly appreciated. The research was funded by the 
Deutsche Forschungsgemeinschaft (DFG) through grant number VE 933/2-1, and 
benefited from the inspiring research environment of the DFG-funded 
Collaborative Research Center CRC1114, "Scaling Cascades in Complex system" 
through the project B07. The exchanges between N.V and D.F were 
greatly facilitated by funding through the DAAD exchange program Procope 
through the project "Data-driven dynamical stability of stably stratified 
turbulent flows". The work of Ivana Stiperski was funded by Austrian Science 
Fund (FWF) grant T781-N32. Davide Faranda acknowledges support from the ERC grant  No. 338965-A2C2. We are grateful to anonymous reviewers for insightful comments that helped us improve the manuscript.

\section*{conflict of interest}
The authors declare no conflict of interests.

\section*{Appendix}
\cite{Lucarini:2016ug} recently proposed a methodology to estimate the 
persistence and local dimension of dynamical states based on timeseries of 
dynamical observables, combining the Poincar\'e recurrence theorem with theory 
of extreme value statistics. In their framework, points on the attractor are 
characterized by parameters of extreme value probability distributions. For a 
given initial point $\zeta$ on a chaotic attractor, the probability of a 
dynamical trajectory $x(t)$ to return within a spherical neighbourhood of the 
initial point has been shown to follow a generalized Pareto distribution 
\citep{MoreiraFreitas:2010cj}, which is a standard distribution in extreme 
value statistics and is a modified exponential law. The time series in this 
context is the distance between $\zeta$ and the other observations along the 
trajectory
\begin{equation}
g\left( x(t)\right) = -\log \left( \delta(x(t), \zeta)\right),
\end{equation}
where $\delta(x,y)$ is a distance function between two points (e.g. the Euclidean distance). Taking the logarithm increases the discrimination of small values of $\delta(x, y)$ and large values of $g(x(t))$ correspond to small distances from the point $\zeta$. The probability of logarithmic returns in the neighbourhood around $\zeta$ can then be expressed as
\begin{equation}\label{Eq:probExp}
P \left(g\left( x(t)\right) > q, \zeta \right) \simeq \exp \left[ -\frac{x-\mu(\zeta)}{\sigma(\zeta)}\right],
\end{equation}
and the parameters of the exponential law $\mu$ and $\sigma$ depend on the point $\zeta$. The local dimension of the dynamics around the point $\zeta$ is finally given by (see \cite{Lucarini:2016ug} for proofs and numerical verifications, and \cite{Faranda:2017bi} for an application to climate dynamics)
\begin{equation} \label{Eq:dim}
d(\zeta) = \frac{1}{\sigma(\zeta)}
\end{equation}
In equation (\ref{Eq:probExp}), $q$ is an exceedance threshold, and is linked 
to the radius $\epsilon$ of the spherical neighbourhood of $\zeta$ via $q = 
g^{-1}(\epsilon) = exp(-\epsilon)$. In other words, requiring the trajectory to 
fall within a sphere around the point $\zeta$ is equivalent to requiring the 
series of $g(x(t))$ to be over the threshold $q$, which can be simply set as a 
percentile of the series itself.\par 
The quantification of the persistence of the state $\zeta$ follows rather 
intuitively: the longer the dynamical trajectory stays in the spherical 
neighbourhood of the point $\zeta$, the more persitent is the dynamics in this 
state. This residence time can be computed by introducing a further parameter 
$\theta$ in the probability distribution (\ref{Eq:probExp}), known as extremal 
index:
\begin{equation}\label{Eq:probExt}
P \left(g\left( x(t)\right) > q, \zeta \right) \simeq \exp \left[ -\theta \left( \frac{x-\mu(\zeta)}{\sigma(\zeta)}\right) \right],
\end{equation}
This parameter $\theta$ can be interpreted as the inverse of the mean residence 
time within the spherical neighbourhood of $\zeta$. From (\ref{Eq:probExt}) it 
follows that $0 < \theta < 1$, where low values correspond to high persistence 
of the trajectory in the neighbourhood of $\zeta$, while values close to 1 
imply that the trajectory immediately leaves the sphere 
\citep{Lucarini:2016ug}.\par

%\printendnotes

% Submissions are not required to reflect the precise reference formatting of the journal (use of italics, bold etc.), however it is important that all key elements of each reference are included.
\clearpage
\pagebreak

\bibliographystyle{plainnat}
\bibliography{bibliography}

\end{document}